\documentclass[12pt]{article}
\usepackage{epsfig}
\usepackage{amsmath,amssymb}
\begin{document}
\vspace*{-.6in} \thispagestyle{empty}
\begin{flushright}
DESY--09--116
\end{flushright}
\baselineskip = 20pt

\vspace{.5in} {\Large
\begin{center}
{\bf On Inflation in the Presence of a Gaugino Condensate}

\end{center}}

\vspace{.5in}

\begin{center}
{\bf  Oleg Lebedev$^{~a}$, Chlo\'e Papineau$^{~a}$ and  
  Marieke Postma$^{~b} $}  \\

\vspace{.5in}

a: \emph{DESY Theory Group, 
Notkestrasse 85, D-22607 Hamburg, Germany
 }
\\ 
b: \emph{NIKHEF, Kruislaan 409, 1098 Amsterdam, The Netherlands  }
\end{center}

\vspace{.5in}

\begin{abstract}
We study the effect of inflation on gaugino condensation in supergravity. Unless the Hubble
scale $H$  is significantly below the gaugino condensation scale, the gaugino condensate 
is a dynamical variable which 
cannot be integrated out. For a sufficiently high $H$, the gaugino 
condensate evolves to zero which in turn leads to dilaton/moduli destabilization. In practice,
this often occurs at the Hubble rate about an order of magnitude below the gaugino 
condensation scale.
This effect is independent of the specifics  of moduli stabilization and thus places model
independent constraints on inflationary scenarios. It also applies more generally to any periods
of fast expansion in the early Universe. 
\end{abstract}

\noindent

\newpage

\section{Introduction}

Gaugino condensation \cite{Veneziano:1982ah}  is arguably  the most attractive mechanism for creating
the hierarchy between the Planck and electroweak (EW)  scales \cite{Nilles:1982ik}-\cite{Dine:1985rz}. 
Starting with a perturbative gauge coupling at the Planck or string scale, a new scale $\Lambda$, 
at which the corresponding gauge group becomes strongly coupled,
is created by dimensional transmutation. If these dynamics  break
supersymmetry, the EW scale can be generated as $\Lambda^3/M_{\rm Pl}^2$
when supersymmetry breaking is communicated by gravity  to the observable sector.
This idea finds support in explicit string models which produce the (exact)  spectrum
of the minimal supersymmetric Standard Model (MSSM) \cite{Lebedev:2006tr}.

If gaugino condensation is indeed part of reality, it must be combined with 
inflation \cite{Guth:1980zm}, which has by now  gathered  strong observational evidence.
The purpose of this paper is to study the dynamics of the gaugino condensate 
with  the background of inflation or, more generally, in the presence of large positive vacuum energy.
Some facets of this problem have been considered before. Gaugino condensation 
generates a potential for moduli, which is modified in the early Universe 
during inflation and   reheating.
To ensure that moduli do not run away, certain constraints on the reheating temperature \cite{Buchmuller:2004xr}
and the Hubble parameter must be satisfied. In the latter case, the situation is
model dependent and only particular (Kachru-Kallosh-Linde-Trivedi-type \cite{Kachru:2003aw})  scenarios
with a specific choice of the inflaton 
 have been studied \cite{Kallosh:2004yh}.
In the present work, we approach this problem from a more general perspective
based on properties of the gaugino condensate itself and without specializing to a 
particular inflationary model or moduli stabilization mechanism. 
  This allows us to obtain a (largely) model independent constraint on the Hubble rate
during inflation or any period of fast expansion.

\section{Veneziano--Yankielowicz potential in supergravity}

In the Veneziano--Yankielowicz approach \cite{Veneziano:1982ah},
gaugino condensation is described in terms of the chiral superfield $U= {\rm Tr}~ W^\alpha W_\alpha$,
with $ W_\alpha $ being the  gauge multiplet  superfield which contains the gaugino as its
lowest component. The low energy effective action for this field is derived using symmetries of 
the system and anomaly cancellation.
In supergravity, the resulting K\"ahler potential ${\cal K}$  and the superpotential $W$
are given by \cite{Burgess:1995aa}
\begin{eqnarray}
&& {\cal K}= -3 \log \Bigl[ e^{-K/3} - a (UU^*)^{1/3}   \Bigr] \;,  \nonumber\\
&& W = {1\over 4}~ U \Bigl ( f S +{2c\over 3} \log (\xi U)   \Bigr) + \tilde W\;. \label{KW1}
\end{eqnarray}
Here we have taken the gauge kinetic function to be given by  the dilaton $S$\footnote{We are 
neglecting threshold corrections to the gauge coupling.}; 
$K$ and $\tilde W$ are the K\"ahler potential   and the superpotential for 
other fields of the system apart from $U$; $a,f,\xi$ are constants of order one
and $c$ is the beta function coefficient 
\begin{equation}
c={3\over 16 \pi^2}~ C(G) \;,
\end{equation}
with $C(G)$ being the quadratic Casimir of the condensing gauge group $G$.

We will be studying   the dynamics of the condensate in the  
region of physical interest  $U \ll 1$ in Planck units.
Changing variables $U \equiv u^3/(3 a )^{3/2}$ and expanding  the result in powers
of $u$, we get
\begin{eqnarray}
&& {\cal K}= K + e^{K/3} u u^* + {1\over 6}~ e^{2 K/3} (u u^*)^2 +... \;,  \nonumber\\
&& W =  u^3 ( S + 2 c \log u ) + \tilde W\;, \label{KW2}
\end{eqnarray}
where, for simplicity, we have set $4 (3a)^{3/2}=1$, $f=1$ and $\xi =(3a)^{3/2} $.

The supergravity scalar potential for this system  is given by 
\begin{equation}
V= e^G (G_i G_{\bar j} G^{i \bar j} -3  ) ~.
\end{equation}
Here the subscript $ l $ ($\bar l$)  denotes differentiation with respect to 
the $l$-th ($l$-th complex conjugate)  scalar field;  
$G$ is a function of the K\"ahler potential ${\cal K}$ and superpotential $W$:
$G={\cal K} +\ln(|W|^2)$, and  $G^{i \bar j}$ is the inverse of $G_{\bar j i}$.

To understand the Veneziano--Yankielowicz result, let us start with   $\tilde W=0$. 
At $u \ll 1$, the dominant contribution to the potential is given by
\begin{equation}
V \simeq e^G \vert G_u \vert^2 G^{uu^*} \;.
\end{equation}
The stationary points of this function are at $W=0, W_u=0$ and $W_{uu}=0$.
The   usual   Veneziano--Yankielowicz solution corresponds to 
$W_u=0$:
\begin{equation}
u_{\rm min}= e^{- {S\over 2c}- {1\over 3}} ~, \label{min}
\end{equation}
which describes a supersymmetric vacuum with massive excitations. 
(More precisely, for an SU$(N)$ group  there are $N$ vacua which 
differ by a phase factor in $u$; this, however, is not important
for our purposes.)
The solution to $W_{uu}=0$ is a local maximum at 
\begin{equation}
u_{\rm max}= e^{- {S\over 2c}- {5\over 6}}~. 
\end{equation}
Finally, $u=0$ formally corresponds
to another supersymmetric chirally invariant vacuum, Fig. \ref{destabplot}.
However, the  Veneziano--Yankielowicz potential cannot be trusted at
$u \rightarrow 0$. The existence of a  supersymmetric   chirally invariant vacuum
is not allowed by general considerations \cite{Cachazo:2002ry} 
 and also inconsistent with the 
Witten index theorem \cite{Witten:1982df} (see also \cite{Goldberg:1995yt}). 
The interpretation of the state at $u=0$ remains 
controversial and it has been conjectured that it corresponds to a non-supersymmetric
(unstable) state \cite{Shifman:2007kt}\footnote{We are grateful to M. Shifman for clarifying this point.}. 
In any case, the   Veneziano--Yankielowicz potential is trustable  around the
SUSY minimum  (\ref{min}) and since the local maximum is close to it,
$u_{\rm max}= u_{\rm min}/\sqrt{e}$, the existence of a potential barrier between
the SUSY vacuum and some other state at $u \ll u_{\rm min}$ is 
also expected to be reliable. 

In the SUSY vacuum (\ref{min}), the gaugino condensate corresponds to a heavy
field and can be integrated out. This creates an effective superpotential for the 
dilaton, which is a necessary ingredient for addressing the problem of dilaton/moduli
stabilization. However, in the early Universe this procedure is not always consistent:
if the expansion rate of the Universe is close to or greater than the gaugino
condensation scale, the condensate remains a dynamical field whose evolution
has to be taken into account. To address this issue, in the next section 
we study the behavior  of the condensate
in the presence of large vacuum energy.

\section{Inclusion of  an inflaton}

Consider the system of the dilaton, gaugino condensate and an extra field $\phi$ 
which generates  large vacuum energy (``inflaton''). This system can be described by
Eq.(\ref{KW1}), and consequently Eq.(\ref{KW2}),    with
\begin{eqnarray}   
&& K= K(S) + K(\phi) \;, \nonumber\\
&& \tilde W= \tilde W (\phi) \;,
\end{eqnarray}
where  $\tilde W (\phi) \gg u^3(S+2c \log u)$.
Since $u \ll 1$, one can expand the scalar potential in powers of $u$. 
Including terms up to forth order, we  get
\begin{eqnarray}
V&=& V_0 + {2\over 3} ~e^{K/3}V_0 ~u u^*    \label{V}   \\ 
&+&
\biggl( e^K \tilde W^* \biggl[ K_{S \bar S}^{-1} K_{\bar S} -{2c\over 3} \Bigl(
  K_{S \bar S}^{-1} \vert K_S \vert^2 + K_{\phi \bar \phi}^{-1} \vert K_\phi \vert^2
+  K_{\phi \bar \phi}^{-1} K_\phi W_\phi^*/ \tilde W^* -3
\Bigr)  \biggr]~ u^3 + {\rm h.c.} \biggr)  \nonumber\\
&+&  e^{2K/3} \biggl( \Bigl\vert u^2 (3 S +6c \log u +2c) \Bigr\vert^2 + {1\over 3}~ V_0 ~(u u^*)^2  \biggr)      
+...               \nonumber
\end{eqnarray}
Here the vacuum energy is given by
\begin{equation} 
V_0= e^K \Bigl(    K_{S \bar S}^{-1} \vert \tilde W K_S   \vert^2 +
 K_{\phi \bar \phi}^{-1} \vert   \tilde W K_\phi + W_\phi    \vert^2 
-3 \vert \tilde W  \vert^2
\Bigr) \;.  \label{V0}
\end{equation}
We see that the condensate receives  mass  of order the Hubble scale ($V_0= 3H^2$) as expected
from general considerations \cite{Dine:1995uk}.
The ${\cal O}(u^3)$ contribution  is an analog of the A--term, while the  ${\cal O}(u^4)$ terms
include  the   Veneziano--Yankielowicz potential $\vert W_u\vert^2$ and an extra contribution
proportional to $V_0$. Note that the potential for the canonically
normalized condensate is obtained by the rescaling $u=e^{-K/6} \tilde u$.

For a sufficiently large $H$, the mass term will dominate and the condensate will quickly evolve 
to zero, $\tilde u \sim e^{-H t} \tilde u_0$. This is intuitively clear since the Hubble expansion is analogous
to ``heating up'' the condensate to temperature of order $H$ (see, e.g. 
\cite{Linde:2005ht}). 
To determine the critical expansion rate, we need to find $V_0$ at which the   Veneziano--Yankielowicz
minimum disappears. 
A sufficient  condition for the absence of local extrema  (apart from $u=0$)
is that the  curvature of the potential in the
$u,u^*$ direction be non-negative, 
\begin{equation}
V_{u \bar u}^2 - V_{uu} V_{\bar u \bar u} \geq 0 \;. \label{pos}
\end{equation}
We are interested in the case when the vacuum energy $V_0$  is dominated by the inflaton F-term,
\begin{equation}
F^\phi \gg F^S \;,
\end{equation}
where $F^i=  e^{G/2} K^{i \bar j} G_{\bar j}  $, which corresponds  to  domination of 
the second term in (\ref{V0}). 
Then the ${\cal O}(u^3)$ contribution in Eq.(\ref{V}) is  (up to a phase)
\begin{equation}
-{2c\over 3} \sqrt{V_0}~ e^{K/2} K_{\phi \bar \phi}^{-1/2} K_\phi ~ u^3 \;. 
\end{equation}
Consider first the case $K_{\phi \bar \phi}^{-1/2} K_\phi \leq {\cal O}(1)$.
Then the cubic term is
suppressed by the loop factor $c$. Further, the  ${\cal O}(u^4)$ term proportional to $V_0$
is small compared to the Veneziano--Yankielowicz piece $\vert W_u\vert^2$   
and can be neglected. 
As a result, the  non-negative curvature    condition 
amounts  approximately to 
\begin{equation}
{2\over 3}~ e^{K/3} V_0 + e^{2K/3} \vert W_{uu}\vert^2 - e^{2K/3} \vert W_{uuu} W_{u} \vert \geq 0 \;.
\end{equation}
Although this inequality  cannot be solved exactly, one can estimate 
the critical $ V_0=3 H^2_{\rm crit}$ by requiring non-negative curvature at the local maximum 
of  the Veneziano--Yankielowicz potential
$u_{\max}$, where
 $ W_{uu}=0$. 
For a canonically normalized $\tilde u = e^{K/6}u $,
we then have 
\begin{equation}
H_{\rm crit} \sim  c~ \vert \tilde u_{\max} \vert \;.
\end{equation} 
This agrees with our numerical results. Note that the cubic term in (\ref{V})  is at most ${\cal O}(cH\tilde u^3)$,
while the quadratic and the  Veneziano--Yankielowicz pieces around $u_{\rm max}$   are ${\cal O}(H^2 \tilde u^2)$ and 
${\cal O}(c^2 \tilde u^4)$, 
respectively, such that for $H> c~ \vert \tilde u_{\rm max} \vert$ the quadratic 
term dominates.

For $K_{\phi \bar \phi}^{-1/2} K_\phi \gg 1$, the cubic term is important and   Eq.(\ref{pos}) gives
\begin{equation}
H_{\rm crit} \sim  c~ \bigl\vert  K_{\phi \bar \phi}^{-1/2} K_\phi~   \tilde u_{\max} \bigr\vert \;.
\end{equation} 

\begin{figure}[htbp]
\begin{center}
\hspace{-1cm}
\includegraphics[scale=1]{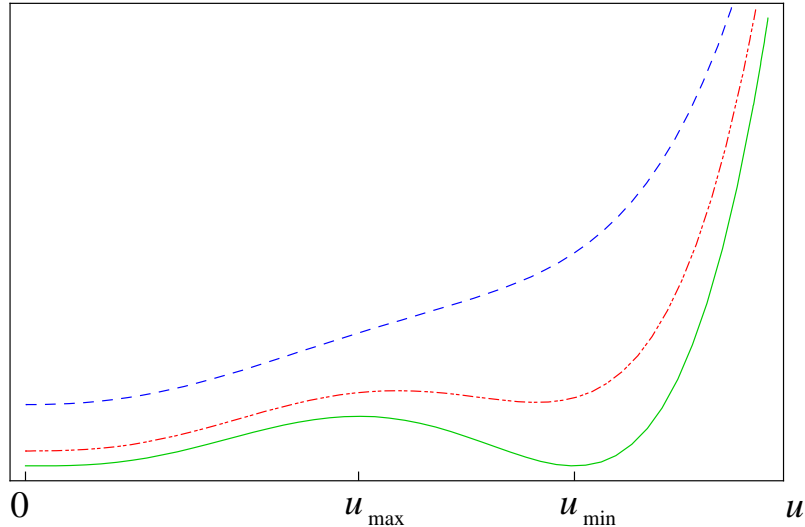}
\caption{The gaugino condensate potential as a function of  vacuum energy. 
The solid (green) curve represents  the Veneziano--Yankielowicz  potential  
($H=0$), while the dotted (red)  and the dashed (blue) curves  correspond to $H < H_{\rm crit}$ and $H \gtrsim H_{\rm crit}$, respectively.\label{destabplot}} 
\end{center}
\end{figure}

The potential for the gaugino condensate in the presence of positive vacuum energy 
is illustrated in Fig. \ref{destabplot}.

\section{Discussion and generalizations}

We  find that, unless $K_{\phi \bar \phi}^{-1/2} K_\phi \gg 1$ , 
the Veneziano--Yankielowicz minimum disappears  at the Hubble rate
a loop factor below the gaugino condensation scale. This is natural as the features 
of the Veneziano--Yankielowicz potential are due to the loop--induced term $2c \log u$
and, consequently, the curvature of the potential at the maximum is loop--suppressed.
In phenomenologically interesting cases, the condensing gauge group is of intermediate
size, e.g. SU(5), such that $c= {\cal O}(10^{-1})$ and the critical Hubble rate is
about an order of magnitude below the  gaugino condensation scale.

This means that for $H>H_{\rm crit}$  the condensate will evolve to zero within 
a few Hubble times. Consequently, the dilaton superpotential approaches a constant
and the dilaton potential attains a run--away form
\begin{equation}
e^K  \times {\rm const}
\end{equation}
with the constant of order $H^2$. The dilaton will thus quickly (within a few Hubble times)    
evolve to weak
coupling $S \gg 1$. Needless to say, this scenario is  phenomenologically  unacceptable
and the constraint $H < H_{\rm crit}$ must be satisfied.\footnote{Note also that as
$H$ decreases,  $u$ will settle in the ``wrong'' vacuum which is separated from
the   Veneziano--Yankielowicz minimum by a large potential barrier.}

This conclusion applies regardless of the specifics of the dilaton stabilization mechanism.
Indeed, the essential ingredient for dilaton stabilization is the superpotential 
due to gaugino condensation which becomes unavailable at $H > H_{\rm crit}$.
Consider, for example, the  K\"ahler stabilization scheme  
\cite{Binetruy:1996xja},\cite{Casas:1996zi}  with
 $K(S) = -\log (S+ \bar S)
 + \Delta K_{\rm np } (S)$. The local minimum in $S$ is obtained due to cancellations
between the K\"ahler corrections and the gaugino condensation superpotential. 
Since the latter is not available as the gaugino condensate  evaporates,
generically no local minima appear during inflation and the dilaton runs away. 
Further, in the racetrack models  \cite{Krasnikov:1987jj},\cite{de Carlos:1992da}  one sums over different condensates in
Eq.(\ref{KW1}):
\begin{eqnarray} 
&& a (UU^*)^{1/3} \rightarrow  \sum_i a_i (U_i U_i^*)^{1/3} \nonumber \\
&& {1\over 4}~ U \Bigl ( f S +{2c\over 3} \log (\xi U)   \Bigr) \rightarrow 
{1\over 4}~  \sum_i U_i \Bigl ( f_i S +{2c_i\over 3} \log (\xi_i U_i)   \Bigr) \;.
\end{eqnarray}
Each one of them attains mass of order $H^2$ during inflation and is destabilized at the Hubble rate
greater than the largest $H_{\rm crit}$ (in fact, a local minimum in $S$ disappears even if only
some of the condensates evaporate). The  potential becomes $\sim  H^2/(S + \bar S)$
and the dilaton runs away.

It is important to note that
our result applies not only to inflation but more generally 
 to any periods of fast expansion in the early Universe. This is  because the slow roll
condition is not essential and the time scale for the evolution of $u$ and $S$
is given by a few Hubble times. Also, an extension to multiple inflatons 
$K(\phi) \rightarrow K(\phi_i), \tilde W(\phi)\rightarrow \tilde W(\phi_i)  $
is straightforward.

Finally, we have taken the gauge kinetic function to be  given by the dilaton.
This can  readily be  generalized to other cases, e.g. in KKLT-type models 
\cite{Kachru:2003aw}
one replaces $S \rightarrow T$ with $T$ being the K\"ahler modulus, 
\begin{eqnarray}
&& K(T)=-3~ \log (T +\bar T)\;, \nonumber\\  
&& W =  u^3 ( T + 2 c \log u )  +  \tilde W   \;,
\end{eqnarray}
where $\tilde W   = W_0 + \tilde W(\phi)$ and 
$W_0$ is the constant superpotential used to stabilize $T$. To have low energy 
supersymmetry, this constant must be adjusted to be very small, ${\cal O}(10^{-13})$.
Again, for large positive vacuum energy, the gaugino condensate acquires mass and
quickly evolves to zero, which leads to disastrous  consequences. 
This happens regardless of the details of the ``uplifting'' mechanism which 
adjusts  the   vacuum energy after inflation.

Let us now discuss our main assumptions.
To establish  evaporation of the gaugino condensate at high $H$, we have relied (1)
on the shape of the  Veneziano--Yankielowicz potential
around the SUSY minimum, which is quite reliable, and (2) on
the K\"ahler potential of the form $(UU^*)^{1/3}$ for small $U$. The latter 
is in fact not necessary and our conclusion would hold  more generally for   
 K\"ahler potentials which can be brought
to the canonical form by a  change of variables $U\rightarrow f(U)$ 
with $f(0)=0$ (and non-singular scalar potential). In this case,
the inflation--induced  mass term is positive and the condensate evolves to zero. 
This  fits the intuitive picture that the  gaugino condensate vanishes
at high de Sitter temperature.

We have also assumed that inflation is driven by the inflaton $\phi$  and the dilaton
 does not play any significant role in it. If this is not the case,
$F^\phi \sim F^S$, the ${\cal O}(u^3)$ term 
proportional to $K_{S \bar S}^{-1} K_{\bar S}$
in Eq.(\ref{V})
becomes important
and  can generate a local minimum at $u>0$ during inflation. Then the gaugino 
condensate will evaporate only for $H \gg u_{\rm max}$. In this case, inflation
does not  amount to  a background for the evolution of the condensate since  
there is significant 
superpotential interaction between the dilaton and the condensate.

Let us remark that realistic vacua with broken
supersymmetry after inflation are possible due to the presence
of extra fields in Eq.\ref{KW1}. In this case, the constraints of
Ref.~\cite{GomezReino:2006wv}  can be satisfied, for example, 
when SUSY breaking is dominated by matter--like fields
\cite{Lebedev:2006qq}.
(Alternatively, one can use non-perturbative corrections to the  
K\"ahler potential \cite{Binetruy:1996xja},\cite{Casas:1996zi}.)
Once a  dilaton stabilization mechanism is employed,
the VY vacuum corresponds to a local minimum in the $S-U$ plane,
so the usual procedure of integrating out the condensate is 
justified. The critical Hubble rate for  a specific model
depends on which field gets destabilized first. 
For instance, $S$ can run away to weak coupling $S \rightarrow \infty$
due to positive vacuum energy.
Thus, during inflation,  the dilaton may be destabilized
before $U$ is, depending on the size of the barrier separating
the local minimum in the $S$ direction from the run-away minimum. 
For instance, in the Kallosh-Linde model \cite{Kallosh:2004yh}  this 
barrier is large, so $ U $ is destabilized before $S$ is, while in the 
usual racetrack model it is   the dilaton that gets destabilized first. 
In any case, the model--independent bound derived in this paper applies regardless
of the moduli stabilization mechanism.

Finally, a comment on loop corrections is in order. In the vacuum, supersymmetry
(and R-symmetry) is broken, so one expects SUSY breaking loop corrections.
These are governed by $ m_{3/2}^2 /16\pi^2 $ and  can only be relevant to 
the dilaton direction, whose mass   is 
$\geq  { \cal O}  (m_{3/2})$ (see e.g. \cite{Buchmuller:2004xr}). 
Such corrections are subleading and 
since the precise
value of the dilaton mass is unimportant for our purposes, we neglect these effects.

\section{Conclusion}

We have studied  the behavior of the gaugino condensate in the presence 
of large vacuum energy. 
If the expansion rate of the Universe is close to or higher than the gaugino condensation
scale, the condensate cannot be integrated out. 
We find that for Hubble rates above a critical value,
the gaugino condensate evolves to zero which leads to dilaton/moduli destabilization.
When the vacuum energy is dominated by the inflaton ($\phi$) 
other than the field $(S)$ producing  the gauge coupling for the condensing gauge group, 
the critical  Hubble rate is given by
\begin{equation}
H_{\rm crit} \sim {\rm max} \{ 1, \bigl\vert K_{\phi \bar \phi}^{-1/2} K_\phi \bigr\vert  \} ~c~  
\vert\tilde u_{\rm max} \vert
\end{equation}
with 
$\tilde u_{\rm max}=  e^{K/6 - S/(2c) - 5/6} $ and $c$ being the one loop beta function
coefficient. Thus, it is typically an order of magnitude below the corresponding
gaugino condensation scale.
This result is independent of the specifics of moduli stabilization
and thus provides a useful constraint on inflationary models.
It  also applies  more generally to any periods of fast expansion in the early Universe.

It would be interesting to further study the cosmological evolution of the dilaton-gaugino condensate system including a background matter or radiation component, to determine which initial conditions lead to dilaton stabilization \cite{inprogress}.\\

{\bf Acknowledgements.} We are grateful to M. Shifman for useful correspondence.
We would also like to thank S. Ramos-S\'anchez for discussions.

\end{document}